\documentclass{PoS}

\pdfoutput=1

\usepackage{color}
\usepackage[normalem]{ulem}
\usepackage{fancybox}
\usepackage{textcomp}
\title{Recent developments for the testing of Cherenkov Telescope Array mirrors and actuators in T\"{u}bingen}

\ShortTitle{Mirror- and Actuatortests for CTA in T\"ubingen}

\author{J.~Dick$^{a}$, A.~Bonardi$^{a,b}$, S.~Bressel$^{a}$, M.~Capasso$^{a}$, \speaker{S.~Diebold}$^{a}$, F.~Eisenkolb$^{a}$, D.~Gottschall$^{a}$, E.~Kendziorra$^{a}$, G.~P\"{u}hlhofer$^{a}$, S.~Renner$^{a}$, A.~Santangelo$^{a}$, T.~Schanz$^{a}$, and C.~Tenzer$^{a}$ for the CTA Consortium\footnote{Full consortium author list at http://cta-observatory.org}\\
E-mail: \email{puelhofer@astro.uni-tuebingen.de} \email{diebold@astro.uni-tuebingen.de}

{\footnotesize
$^{a}$ Institut f{\"u}r Astronomie und Astrophysik, Abteilung Hochenergieastrophysik, Kepler Center for Astro and Particle Physics, Eberhard Karls Universit{\"a}t, Sand 1, D 72076 T{\"u}bingen, Germany;\\
$^{b}$ now at Department of Astrophysics, Radboud University Nijmegen, Huyge Heijendaalseweg 135, 6525 AJ Nijmegen, The Netherlands}
}
\abstract{The Cherenkov Telescope Array (CTA) is the next generation Cherenkov telescope facility. It will consist of a large number of segmented-mirror telescopes of three different diameters, placed in two locations, one in the northern and one in the southern hemisphere, thus covering the whole sky. The total number of mirror tiles will be on the order of 10,000, corresponding to a reflective area of $\sim$10$^4$\,m$^2$. The Institute for Astronomy and Astrophysics in T\"{u}bingen (IAAT) is currently developing mirror control alignment mechanics, electronics, and software optimized for the medium sized telescopes. In addition, IAAT is participating in the CTA mirror prototype testing. In this paper we present the status of the current developments, the main results of recent tests, and plans for the production phase of the mirror control system. We also briefly present the T\"{u}bingen facility for mirror testing.}

\FullConference{The 34$^{th}$ International Cosmic Ray Conference,\\
		30 July--6 August, 2015\\
		The Hague, The Netherlands}

\begin{document}

\section{Introduction}
The Cherenkov Telescope Array (CTA) is the next-generation ground-based Cherenkov telescope facility currently under preparation by an international consortium \cite{bib:Design2011}. CTA will host on the order of hundred Imaging Atmospheric Cherenkov Telescopes (IACTs) of different sizes, to cover the photon energy range from few tens of GeV to few hundreds of TeV. Three different telescope classes are foreseen: large size telescopes (LSTs) with $\sim$23\,m dish diameter, medium size telescopes (MSTs) with $\sim$12\,m, and small size telescopes (SSTs) with $\sim$4\,m. The reflective surface of each telescope's primary mirror will consist of hexagonal mirror facets of $\sim$0.5-2\,m$^2$ area, according to the telescope type \cite{bib:StatusOfTech}. The total reflective surface of the entire CTA is on the order of $\sim$10$^4$\,m$^2$, hence several thousand mirror segments will be required and high efficiency and throughput during testing, quality control, and alignment of the segments are necessary. In this paper, activities conducted by the Institute for Astronomy and Astrophysics in T\"{u}bingen (IAAT) in the fields of mirror testing and actuator development are reported. 

\section{Mirror reflectance and point spread function testing}

\begin{figure}[t]
\begin{center}
    \includegraphics[width=.65\textwidth]{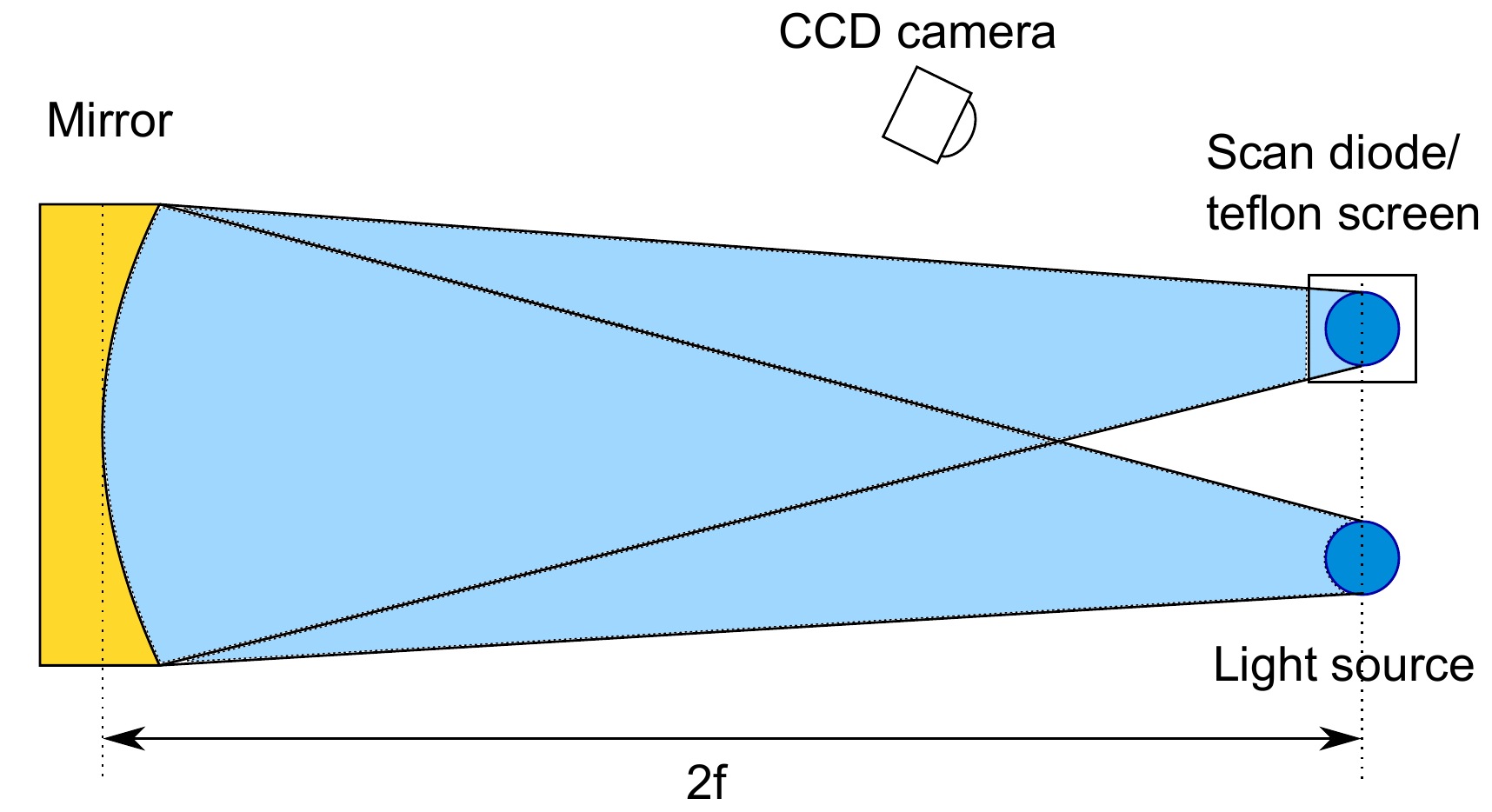}
\end{center}
\caption{Schematic of the 2f setup: An image of the light source, placed at 
    a distance of two times the nominal focal length, is projected on a screen 
    placed at the same distance next to the light source. At the position of 
    the image a scan with a photodiode is performed and an image with a non-UV sensitive CCD camera is taken. In the upgraded setup the CCD camera is replaced by a UV-sensitive model and the scan is no longer performed.}
\label{Figure:2fsketch}
\end{figure}

The mirror segments for IACTs have to fulfill two basic optical performance requirements. The point spread function (PSF) of the mirror should be smaller than a certain limit -- for photo-multiplier-tube cameras the limit corresponds to a fraction of the pixel size in the focus of the telescope -- and the reflectivity of the mirror surface for the typical wavelength range of Cherenkov radiation has to be above a certain threshold. To characterize the prototype mirrors for CTA, two different procedures are followed in T\"{u}bingen:\\
\textit{1)} In order to assess the (undirected) surface reflectivity of the entire mirror, measurements are taken directly on the surface of the mirror at 20-30 different spots (depending on the mirror size) using a commercial handheld spectrometer provided by the Max-Planck-Institute f{\"u}r Kernphysik in Heidelberg. These measurements involve significant manual activity and are quite time consuming ($\geq$1\,h/mirror) but very precise and provide a high spectral resolution.\\
\textit{2)} In a setup which is in an evolved form also planned to be used for mass testing of CTA mirrors, the mirror is placed on a movable support at a distance of two times the nominal focal length of the mirror (2f) to a light source (see Figure~\ref{Figure:2fsketch}). For an ideally focusing mirror, an exact image of the light source is thus projected back in an image plane, e.g. on a screen placed next to the light source. In the current setup, a halogen lamp with an aperture smaller than the expected PSF of the mirror ($\sim$3\,mm) is used.

The measuring device is a photodiode in combination with a filter-wheel connected to a lock-in amplifier and mounted on a movable scanning table. The position of the diode follows a fine grid pattern centered on the reflected light spot in order to record the reflected spatial intensity distribution of the spot. From this scan result, an approximation of the wavelength-depending PSF of the mirror can be calculated. In order to measure the averaged PSF of the mirror, a non-UV-sensitive CCD-camera is also used to take pictures of the spot on the screen, since the CCD's angular resolution is much better than the pixelation of the grid with the photodiode.

To measure the reflectivity, the same setup with the photodiode is used. To minimize systematic errors in the absolute calibration, regular reference measurements are taken by placing the same photodiode at a predefined position between light source and mirror. By integrating the total flux within a predefined circle and dividing this value by the flux at the position of the mirror surface, a value for the (directed) reflectivity is calculated. By using four different filters in a filter wheel placed in front of the photodiode the reflectivity is obtained at the four wavelengths 310\,nm, 400\,nm, 470\,nm, and 550\,nm. This setup has been designed, built, and operated for H.E.S.S. \cite{bib:HESSCrab2006} by the Max-Planck-Institut f\"{u}r Kernphysik in Heidelberg in the past. For the testing of the $\sim 1000$ mirrors of the 28\,m telescope of H.E.S.S. phase II \cite{bib:VincentHESSTwo2006}, the setup was relocated to T{\"u}bingen and has been operated since then at IAAT. The long (100~m) and dark basement of the IAAT building is a suitable location to perform the 2f measurements also for all types of spherical CTA mirror segments.

The current 2f-setup is at the moment undergoing some refurbishments to make it more stable and adapt it better to the setup geometry imposed by the set of CTA mirror focal lengths. This includes replacing the filters with a set of higher transmission (addressing specifically the UV range), and improving the alignment of the light source. As indicated by ongoing tests, the setup is expected to provide reliable results for CTA both on the PSF and the reflectivity of the mirrors with a completely automatically operated setup. The major disadvantage is that even this fully automated measurement for one mirror takes six to eight hours, therefore it cannot be used in the current configuration during the full testing campaign for CTA. 

To overcome this limitation, the IAAT has been investigating a new setup, using the same optical configurations. Instead of working with a grid scan with a photodiode, a high-resolution UV-sensitive ALTA U47 CCD camera is used that takes an image of the light spot reflected by the mirror on a Teflon screen. The new screen has good diffuse reflectivity also in the UV. From this image the size of the light spot is derived, allowing a wavelength-dependent measurement with high angular resolution of the PSF of the mirror. LEDs of different wavelengths are used as a light source to replace the individual filter measurements.

In the reflectivity measurements, the CCD camera is now used at two different positions. The first position is such that the Teflon screen placed between LEDs and mirror can be monitored. This provides a method to obtain a reference flux measurement, similar to the reference measurements of the old setup. The second position serves to image the reflected spot. For the light spot of the mirror, again the intensity inside a predefined circle is calculated and a comparison of this value to the reference measurement yields the reflectivity. In this way, both the PSF and the reflectivity of the mirror can be measured for one wavelength within minutes. Currently, the setup is developed further to allow a higher degree of automatization and to reduce systematic effects emerging from errors of the distance measurement of the LED and the reference position of the screen or from the (very low) remaining non-constant background light in the institute's basement. The final goal after the ongoing upgrade is to provide a complete set of PSF and reflectivity measurements at four different wavelengths in less than an hour, including mounting and un-mounting of the mirror.

\section{Mirror alignment system}

\begin{figure}[t]
\begin{center}
    \includegraphics[width=.50\textwidth]{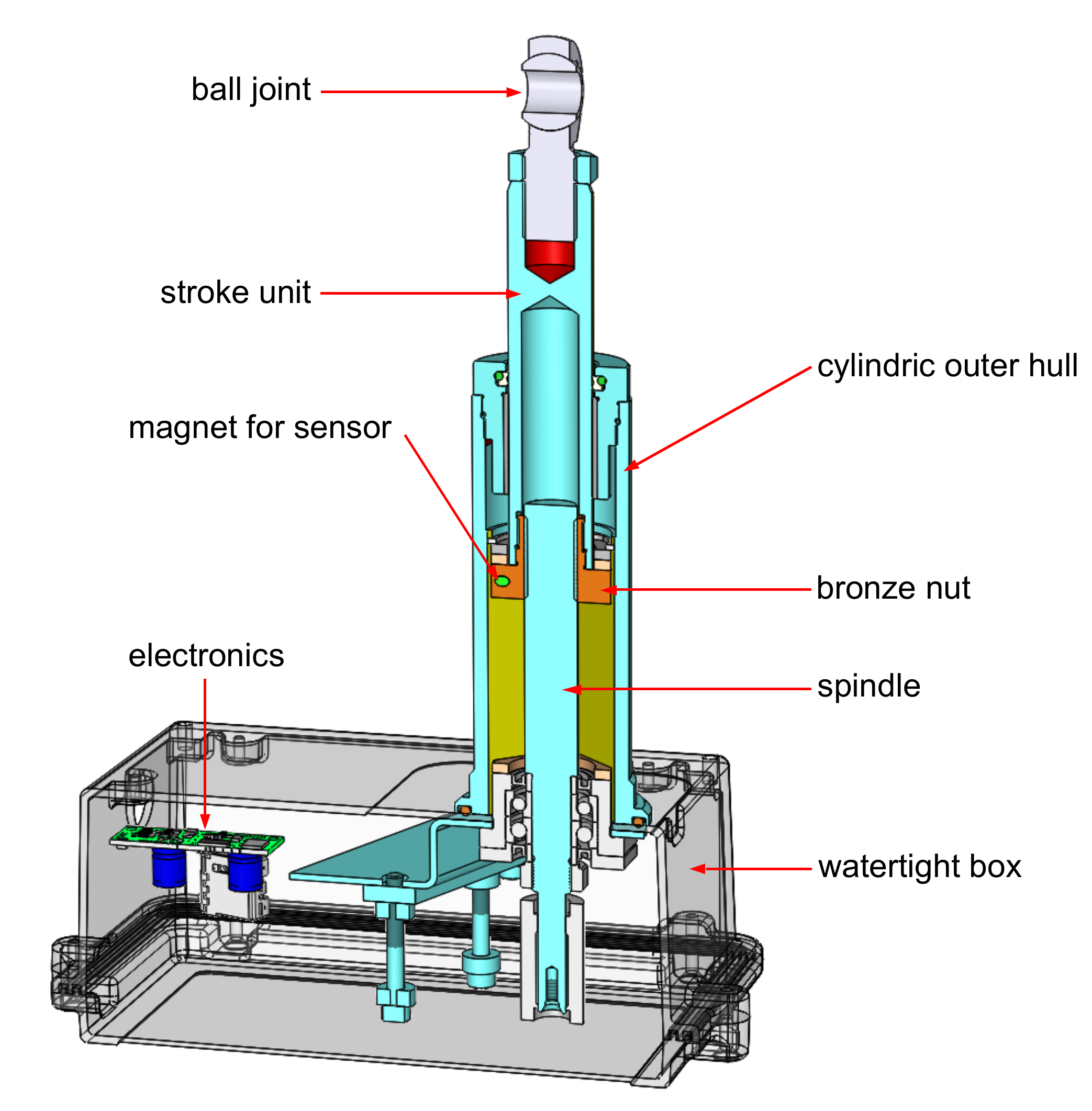}\end{center}
\caption{Current third generation actuator design. A motor (not shown) located in a sealed watertight box is turning the spindle. This results in a lateral movement of the nut, which is connected to the stroke unit. The position of the stroke unit is determined relative to a Reed contact zero point by means of a Hall sensor counting half rotations of the motor.}
\label{Figure:ActuatorCAD}
\end{figure}

The large number of mirror facets on the CTA telescopes, especially for the MST and LST types, requires motorized mirror actuators. This allows to realign the mirror facets easily, e.g. in order to compensate telescope structure deformation at different zenith angles. It is in principle possible to regularly optimize the alignment in order to obtain the best possible telescope PSF. However, a frequent realignment during observation time will probably only be adopted for the LST telescopes. The strain on mirror actuators is certainly lower in the case of occasional alignment procedures. Still options which might be suited for high and low duty cycles are being pursued. The design which the IAAT is currently developing and testing originates in the actuator mechanics developed for the large, 28\,m telescope of H.E.S.S. phase II \cite{bib:HESSSteuerelektronik}. Starting in 2012, IAAT collaborated with Buck Engineering Consulting GmbH to improve the actuator mechanics design with the main goal of optimizing the cost-performance balance. In this context, 32 new pre-series actuators had been produced, installed, and tested for a year on the MST prototype built by DESY in Berlin-Adlershof. With the results from these and other tests conducted at IAAT, we went through a redesign of the concept in collaboration with the company Lesatech in order to also comply with the most recent requirements on earthquake-, ice-, and wind-load stability for all possible telescope sites including Chile. The current actuator design is shown in Figure~\ref{Figure:ActuatorCAD}.

As in the case of the H.E.S.S. 28\,m telescope actuators, mirror facets are supported by two motor-driven actuators and one freely-tilting bolt, which are fixed to the telescope structure. In the current MST design, the actuators are connected through a supporting triangle which is fixed to the structure of the telescope dish. One actuator is tightly anchored to the supporting triangle so as not to allow any tilting movement; the second one is equipped with a pivot, allowing free rotation along a tilting angle. The motor of each actuator is housed in a watertight box vented by a sintered bronze valve. The minimum step size of our actuator system is about 4.8\,\textmu m and the total extension range is $\sim$50\,mm. Furthermore, as in the case of the H.E.S.S. 28\,m telescope actuators, the external fixing allows to shift the actuator position with reference to the triangle structure by about 4\,cm.

Actuators are driven by low-price brushed 12\,V DC motors used in large numbers in the automotive industry (GMPD 404.905 from Valeo). Supply voltage along the cables is 24\,V and -- to fit the motor specifications -- PWM (Pulse Width Modulation) is used, reducing significantly the current and the power dissipation at the cables compared to 12\,V operation. The design includes the possibility to ramp the actuator speed up or down and to have automatic active mechanical hysteresis compensation as well. Positioning is performed in an incremental way by counting signals provided by a Hall sensor that counts half rotations (called 'steps') inside the motor; a relative zero position is determined by a robust Reed proximity switch inside the actuator cylinder. All actuator movements are measured in steps, with one step referring to one count of the motor Hall sensor and 420 steps corresponding to a full rotation of the spindle. As the motor turns the spindle, the nut is moved and carries the stroke unit in and out of the actuator.

\subsection{Actuator control electronics}

\begin{figure}[t]
\begin{center}
     \includegraphics[width=.49\textwidth]{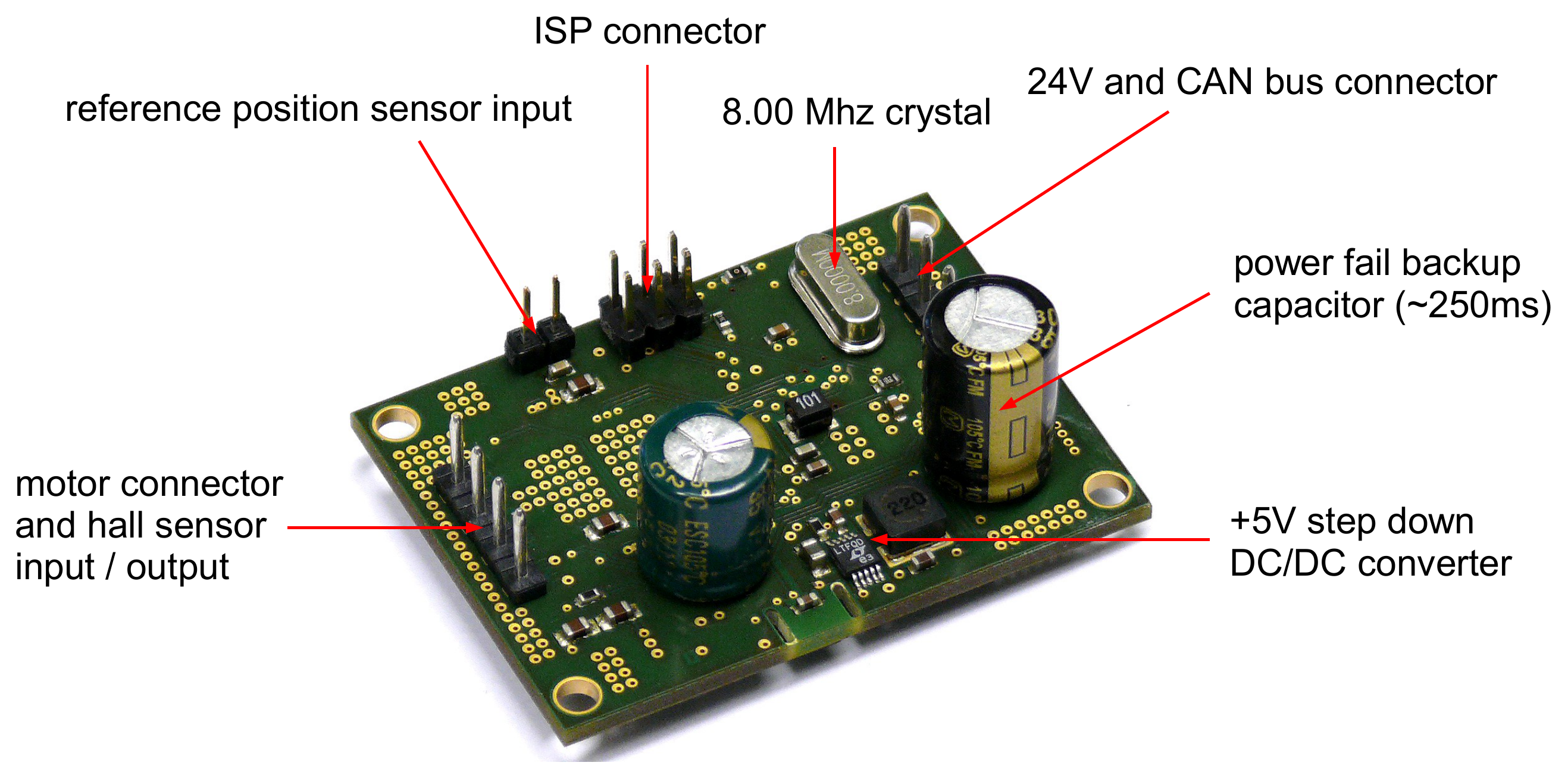}  \includegraphics[width=.49\textwidth]{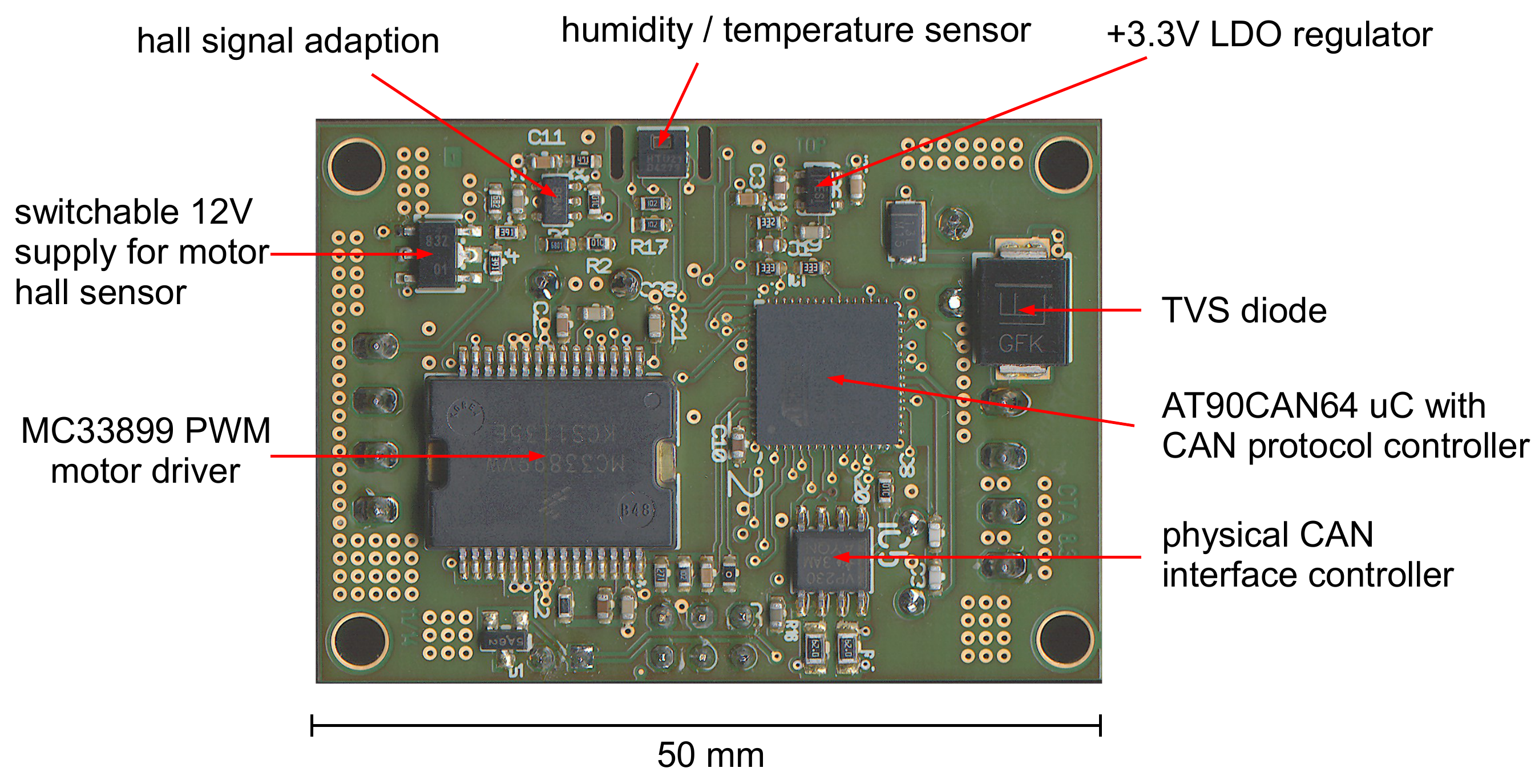}
\end{center}
\caption{Top (left) and bottom view (right) of the current third generation of the actuator control electronics located inside the watertight actuator box.}
\label{Figure:ActuatorElectronics}
\end{figure}

Compared to the electronics which are used to control the actuators at the 28\,m telescope of H.E.S.S. II \cite{bib:HESSSteuerelektronik}, a new architecture based on the Controller Area Network (CAN) interface \cite{bib:CAN_Reference} has been developed and employed. The new system is fast and provides high transmission reliability even for larger distances through the use of differential data lines. Control via CAN requires just two wires in addition to the motor power lines in the cabling of the actuators. Each actuator is representing a single CAN device with unique 11 bit CAN identifier (protocol layer A). A hybrid solution between parallel and serial distribution is used, allowing a flexible number of outgoing CAN branches (up to 16) which can be adapted to any telescope size. Theoretically, more than 2000 actuators can be controlled and run at the same time, only limited by the voltage drop at the power lines. Furthermore, the system is extremely simple and robust as evidenced by our laboratory and outdoor tests. During different tests, millions of CAN commands were transmitted without experiencing any transmission failure. In each actuator box there is an electronic control board (see Figure~\ref{Figure:ActuatorElectronics}), composed of a driver for the motor and a RISC micro-controller (AT90CAN64 with embedded CAN protocol controller) running a multi-threaded real time operating system. Operation parameters are monitored continuously and can be checked at any time by command (e.g. status and position of the actuator, motor current, temperature, and humidity inside the box).  A central controller box serves one telescope, providing a gateway functionality between Ethernet and the CAN buses. For the software interface the goal is to use an OPC-UA server running on the ARM-based industrial grade embedded Linux PC of this center box.  

\subsection{Actuator test results}

\begin{figure}[t]
\begin{center}
     \includegraphics[width=.49\textwidth]{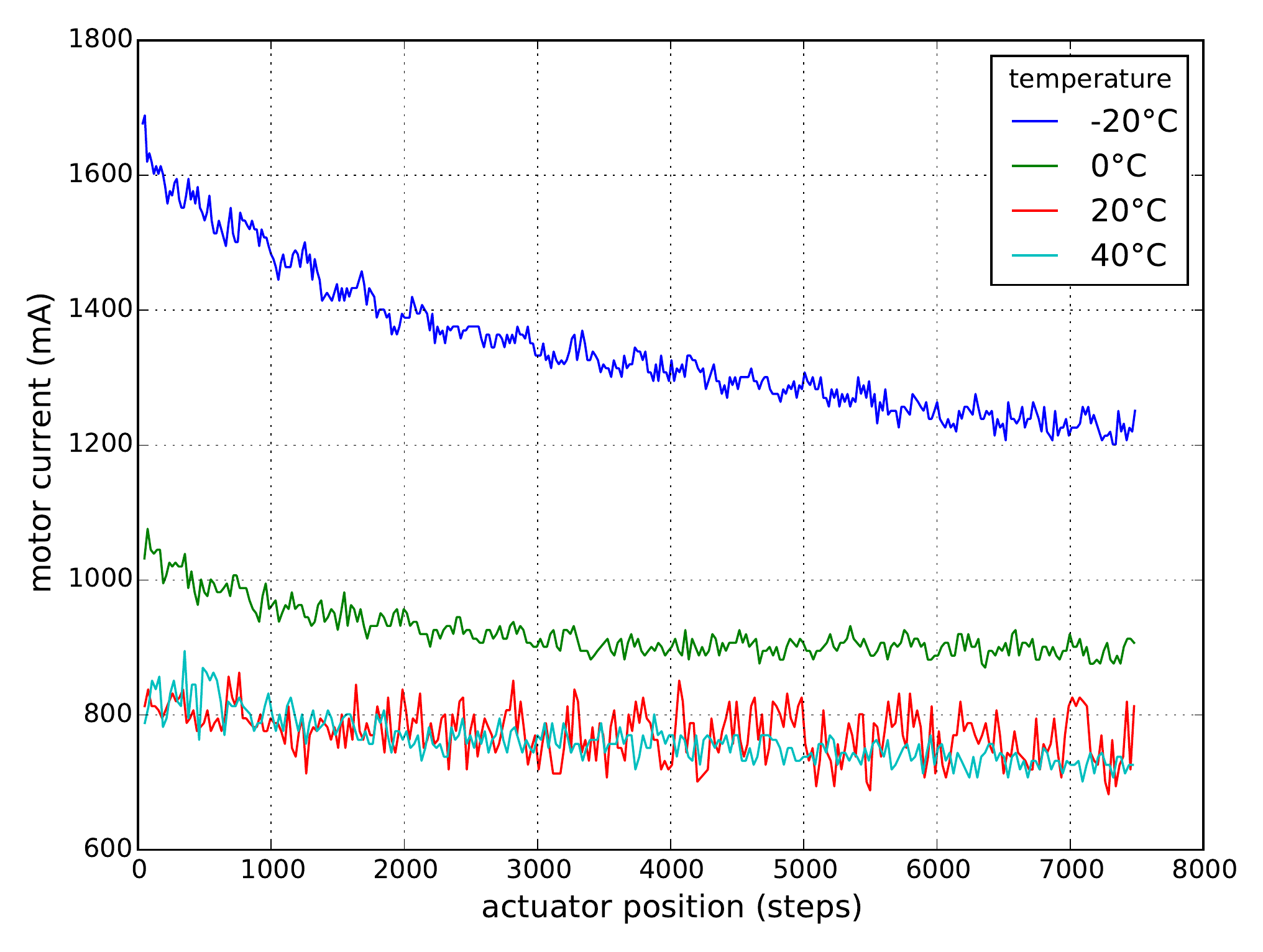}  \includegraphics[width=.49\textwidth]{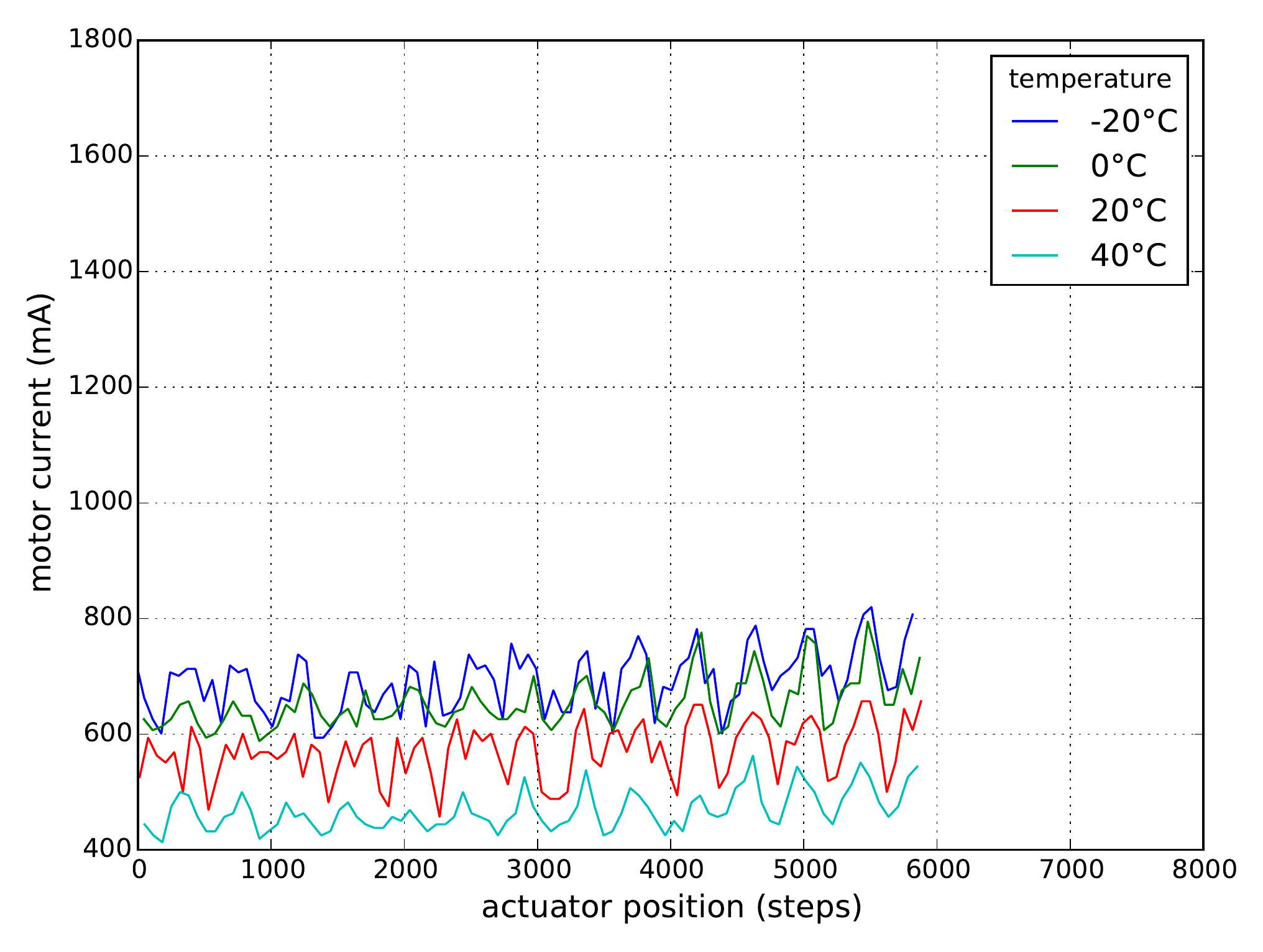}
\end{center}
\caption{Motor current as a function of ambient temperature. Comparison between the second (left panel) and third generation (right panel) mechanics design.}
\label{Figure:TempCurrent}
\end{figure}

\textit{Ambient temperature dependence:} Several long- and short-term load-tests have been carried out with all three mechanics generations of our actuator design so far (the H.E.S.S. 28\,m telescope design, the Buck prototypes for CTA, and the current redesign developed by Lesatech). One key example are temperature dependency tests of the motor current which give an indication of the force necessary to turn the spindle and make the nut move. This is directly related to the manufacturing and material quality and is strongly influenced by the difference of the thermal expansion coefficients of the two involved materials. In earlier design stages, extreme temperatures as well as not moving the actuator for more than a year could get the nut stuck on the spindle and require currents above the shut-off threshold to unlock the actuator. As the motor current values are also measured to check if the motor has reached its outermost position and cannot move further, the fluctuations during operation must be known. Therefore, all actuator designs were tested for operability in a climate chamber and the dependency of the motor current on the temperature was measured in a temperature range from -20\textdegree C to +40\textdegree C under a load of 10\,kg (typical mirror load). Figure~\ref{Figure:TempCurrent} shows the  results for moving upwards against the load. For low temperatures, the motor current is larger at the beginning but then the spindle and nut quickly heat up due to friction. In the current design the Iglidur nut was subsequently exchanged by a bronze nut which has a thermal expansion coefficient better matching the one of the spindle and is lubrication-free, too.

\begin{figure}
\begin{center}
     \includegraphics[width=.52\textwidth]{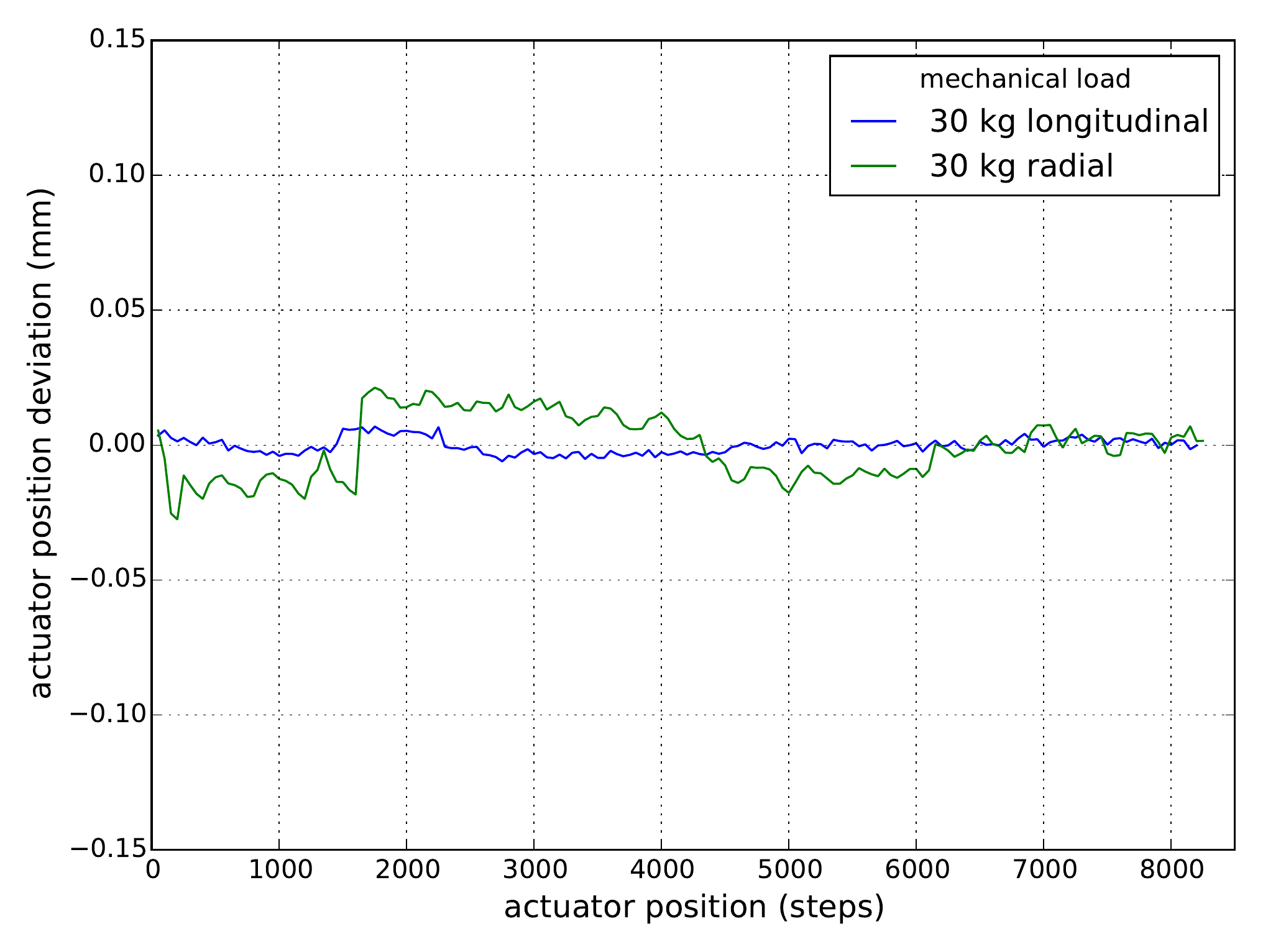}  \includegraphics[width=.47\textwidth]{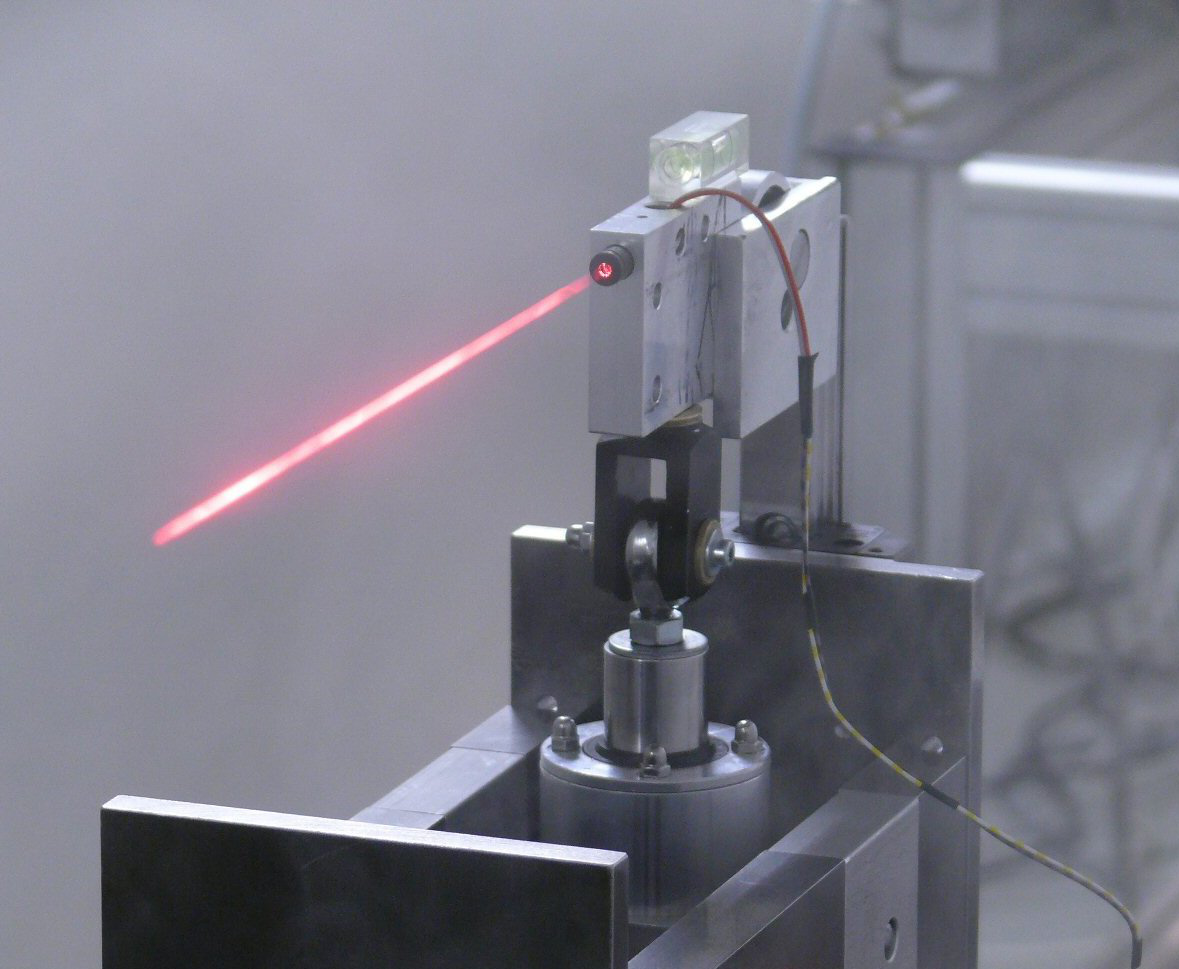}
\end{center}
\caption{Left: Deviation of the measured actuator position (here second generation mechanics) from the calculated one in a test with 30\,kg load pulling at along the actuator axis (blue) and at a right angle to the actuator axis (green). Right: Setup with the laser indicating the current actuator position for the accuracy measurement.}
\label{Figure:LinearDeviation}
\end{figure}

\textit{Long-term tests:} A long duration test has been going on at the IAAT with two actuators of the H.E.S.S. design being constantly outdoors and driven every three minutes, adding up to $\sim$80,000 movements up to now. The second generation actuator prototypes had been mounted on the MST telescope prototype in Adlershof for most of 2013 and 2014 in a long-term test. As a result of these studies, one of the actuators in Adlershof was later found to have had a significant amount of water inside which had shortened the motor and corroded the internal mechanical parts. The sealing concept of the box was consequently rethought and optimized in the current design.

\textit{Positioning accuracy:} The steepness of the spindle thread results in a 2\,mm movement of the actuator for one complete rotation (420 motor steps). This leads to a value of 4.762\,\textmu m/step. The steepness was chosen to meet the resolution goal of better than 20\,\textmu m, but also to provide enough friction to remain in place when there is no motor current present (self-locking mechanism). In order to cross-check the positioning accuracy of the actuator, tests were run to correlate the number of driven steps with the actual actuator position. In the setup a laser was tilted by the actuator movement and the spot of the laser was recorded at a distance of 55\,m. The long distance amplifies the small movement of the actuator and makes the tilt-angle measurable. From the distance of the laser to the target screen and the movement of the laser spot, a triangle is defined from which the actual movement of the actuator stroke unit can be derived. Figure~\ref{Figure:LinearDeviation} shows the setup for this positioning accuracy measurement.

\textit{Bending and tilting:} In addition, the second generation actuators were tested for bending and tilting. During the tests, the actuator was fixed on a table with a granite tabletop. For different positions of the stroke unit, different weights were applied step by step to its tip. At the same time the positions of the outer cylinder hull and the tip of the stroke unit were measured with a dial gauge. For the outer cylinder no change in the position was observed. The observed behavior of the stroke unit has (among other reasons) led to a redesign of the outer cylinder and the stroke unit in order to provide more stability against bending under load. The first tests with the current actuator design show that the tolerance range in the manufacturing of the movable parts is smaller, and therefore there are less intrinsic possibilities for bending. For the outer cylindric hull, a different material is now used in order to provide more guidance and stability for the stroke unit. The first tests with the new design are scheduled in the next months.

\section{Conclusion}

The mirror test facility at IAAT is currently being improved to match better the CTA demands. This is scheduled to be completed in time for the planned mass production of CTA spherical mirror tiles, a fraction of which could be tested at the IAAT facility. IAAT is also developing mirror actuators, specifically tuned for CTA Davies-Cotton MST requirements. The actuator mechanics and electronics are currently revised to better adapt them to the environmental conditions expected at all current CTA candidate sites. For continued testing, it is planned to equip a fraction of the mirrors at the MST prototype telescope in Berlin-Adlershof with the new actuator revision. At the same time, plannings for future mass production are ongoing.

\subsection*{Acknowledgements}

We gratefully acknowledge support from the agencies and organizations under Funding Agencies at www.cta-observatory.org.

\end{document}